\documentclass[11pt]{article}
\pdfoutput=1
\usepackage{jheppub}

\usepackage[svgnames,table]{xcolor}
\usepackage{graphicx}
\usepackage{amsfonts,amsmath,amssymb,amsthm}
\usepackage{mathrsfs,bm,bbm,esint}
\usepackage[mathscr]{eucal}
\usepackage{physics}
\usepackage{slashed,soul}
\usepackage{rotating,pdflscape}
\usepackage{enumerate}
\usepackage{multirow,array}
\usepackage[numbers,sort&compress]{natbib}
\usepackage{tikz}
\usepackage{tkz-euclide}
\usetikzlibrary{cd}
\usetikzlibrary{decorations.pathmorphing}
\usetikzlibrary{decorations.pathreplacing}
\usetikzlibrary{decorations.markings}
\tikzset{snake it/.style={decorate, decoration=snake}}
\tikzset{->-/.style={decoration={
  markings,
  mark=at position .5 with {\arrow{>}}},postaction={decorate}}}
\tikzset{-<-/.style={decoration={
  markings,
  mark=at position .5 with {\arrow{<}}},postaction={decorate}}}
\tikzset{cross/.style={cross out, draw=black, fill=none, minimum size=2*(#1-\pgflinewidth), inner sep=0pt, outer sep=0pt}, cross/.default={2pt}}
\usepackage{subcaption}
\usepackage{float}
\usepackage{afterpage}
\renewcommand{\thefigure}{\arabic{figure}}

\usepackage{cleveref}
\crefname{section}{§\!\!}{§§\!\!}
\Crefname{section}{§}{§§}
\crefname{appendix}{Appendix}{Appendices\!}
\crefname{figure}{Fig.\!}{Figs.\!}

\usepackage{comment}
\usepackage{mathtools}
\usepackage{upgreek}
\usepackage{empheq} 
\usepackage[many]{tcolorbox}

\newtheorem{theorem}{Theorem}[section]

\tcolorboxenvironment{colortheorem}{
  colback=blue!5!white,
  boxrule=0pt,
  boxsep=1pt,
  left=2pt,right=2pt,top=2pt,bottom=2pt,
  oversize=2pt,
  sharp corners,
  before skip=\topsep,
  after skip=\topsep,
}

\newtheorem{definition}{Definition}[section]

\theoremstyle{definition}
\newtheorem{remark}{Remark}[section]

\theoremstyle{definition}
\newtheorem{example}{Example}[section]

\definecolor{rust}{rgb}{0.8,0.2,0.2}

\newcommand{\prn}[1]{\left ( #1 \right )}
\newcommand{\prnbig}[1]{\Big( #1  \Big)}
\newcommand{\brk}[1]{\left [ #1 \right ]}
\newcommand{\brkbig}[1]{\Big [ #1  \Big]}
\newcommand{\cbrk}[1]{\left\{ #1 \right \} }

\newcommand{\md}{\mathbf{m}}
\newcommand{\zm}{\mathcal{Z}_{\mathbf{m}}}

\newcommand{\wm}{\mathcal{W}_{\mathbf{m}}}
\newcommand{\bpts}{\bm{z}}
\newcommand{\perms}{\bm{\sigma}}
\newcommand{\cpone}{\mathbb{CP}^1}
\newcommand{\psimdi}{\psi^{I}_\md}
\newcommand{\piz}{p^I_z}
\newcommand{\cyc}{\gamma}
\newcommand{\pf}{\mathscr{Z}}
\newcommand{\antiholo}{\prn{\text{anti-holomorphic}}}
\newcommand{\tonept}{\expval{T(z)}_\md}
\newcommand{\hurwitz}{\mathcal{H}_{g,N}\prn{\vb*{\lambda}}}
\newcommand{\Hurwitz}{\mathcal{H}_{g}\prn{\vb*{\lambda}}}
\newcommand{\Norb}{\mathsf{N}}
\newcommand{\renyi}{R\'{e}nyi }
\newcommand{\ith}{I^{\text{th}}}
\newcommand{\cft}{\mathcal{C}}

\title{Twist operator correlator revisited and tau function on Hurwitz space}

\author{Hewei Frederic Jia}
\affiliation{Center for Quantum Mathematics and Physics (QMAP)\\
Department of Physics and Astronomy, University of California, Davis, CA 95616 USA}

\emailAdd{fjia@ucdavis.edu}

\abstract{Correlation function of twist operators is a natural quantity of interest in two-dimensional conformal field theory (2d CFT) and finds relevance in various physical contexts. For computing twist operator correlators associated with generic branched covers of genus zero and one, we present a generalization of the conventional stress-tensor method to encompass generic 2d CFTs without relying on any free field realization. This is achieved by employing a generalization of the argument of Calabrese-Cardy in the cyclic genus zero case. The generalized stress-tensor method reveals a compelling relation between the twist operator correlator and the tau function on Hurwitz space, the moduli space of branched covers, of Kokotov-Korotkin. This stems from the close relation between stress-tensor one-point function and Bergman projective connection of branched cover. The tau function on Hurwitz space is in turn related to the more general isomonodromic tau function, and this chain of correspondence thus relates the twist operator correlator to a canonical algebro-geometric object and endows it with an integrable system interpretation. Conversely, the tau function on Hurwitz space essentially admits a CFT interpretation as the holomorphic part of the twist operator correlator of $c=1$ free boson.}
\preprint{}
\newpage

\begin{document}
\maketitle


\section{Introduction and summary of results}
The twist operator correlator is a natural quantity of interest in two-dimensional conformal field theory (2d CFT) due to the well-known relation between 2d CFT and Riemann surfaces and the classical fact that a Riemann surface $\Sigma$ can generally be realized as a branched cover of $\mathbb{CP}^1$ via a meromorphic function $\phi: \Sigma \to \mathbb{CP}^1$. Twist operator correlator admits a path integral definition~\cite{Lunin:2000yv,Calabrese:2004eu}:
\begin{definition}[Twist operator correlator]
\label{def-partition-function}
A monodromy data is a pair $\mathbf{m} = (\bm{\sigma},\bm{z}) \in S^M_N \times \mathbb{C}^M$. Twist operator correlator/partition function with prescribed monodromy $\mathbf{m}$ for a generic 2d CFT $\mathcal{C}$ is defined by path integral for $N$ copies of $\mathcal{C}$ with monodromy conditions for fundamental fields $\{\varphi_I\}_{I = 1, \cdots, N}$ specified by $\bf{m}$
\begin{equation}
    \mathcal{Z}_{\mathbf{m}}(\bm{z}|\bm{\sigma}) = \expval{\prod_i \sigma_i (z_i)} \coloneqq \int\limits_{\varphi_I \prn{\xi_i \circ z}  = \varphi_{\sigma_i(I)}(z)} \brk{D\varphi} e^{-\sum_I S[\varphi_I]}
\end{equation}
where $\bm{\xi} $ are generating loops in $\pi_1 \prn{\mathbb{CP}^1 \setminus \bm{z}}$ and $\xi_i \circ z$ denotes continuation along path $\xi_i$. The monodromy data $\mathbf{m}$ is naturally identified as the monodromy data of a branched cover 
\begin{equation}
    \phi_\md: \Sigma \to \mathbb{CP}^1 \nonumber
\end{equation}
with branch locus $\bpts$ (i.e., critical values of $\phi_\md$) and corresponding permutation monodromies $\perms$. In other words, the twist operator correlator $\zm$ is the partition function of $\cft$ on $\Sigma$ evaluated in the conformal frame where base $\cpone$ has flat metric.\footnote{Technically, a cut-off is required at infinity on $\cpone$; this gives trivial contribution to the $\bpts$-dependence of the twist operator correlator $\zm$~\cite{Lunin:2000yv}. }
\end{definition}

Besides the formal reason mentioned above that justifies twist operator correlator as a natural quantity of interest in 2d CFT, there is indeed a rich literature on the quantity with motivations from various different physical contexts, such as orbifold CFT with motivation from string theory~\cite{Dixon:1986qv,Knizhnik:1987xp,Arutyunov:1997gt}, replica trick calculation of quantum information-theoretic quantities (e.g., entanglement entropy~\cite{Calabrese:2004eu,Calabrese:2009ez,Hartman:2013mia}, entanglement negativity~\cite{Calabrese:2012ew}, reflected entropy~\cite{Dutta:2019gen}, etc.), conformal bootstrap~\cite{Cardy:2017qhl,Cho:2017fzo}, symmetric product orbifold~\cite{Lunin:2000yv,Pakman:2009zz,Dei:2019iym} and the associated $\text{AdS}_3/\text{CFT}_2$~\cite{Eberhardt:2019ywk,Gaberdiel:2020ycd}. For example, in the context of replica calculation of information-theoretic quantities, twist operator correlator with pairwise trivial monodromies admits density matrix interpretation. Famously, the twist operator correlator corresponding to $M=2$ monodromy data $\bf{m}$ with $\bm{\sigma}$
\begin{equation}
    \sigma_1 = \sigma^{-1}_2 = (1\dots N)
\end{equation}
is related to the universal single interval \renyi entropy of ground state of CFT $\mathcal{C}$~\cite{Calabrese:2004eu}:
\begin{equation}
    \mathcal{Z}_{\mathbf{m}} = \Tr\prn{\rho^N_A} = z^{-2h}_{12} \Bar{z}_{12}^{-2\Bar{h}}, \ \ A = [z_1, z_2], \ \ h=\Bar{h}=\frac{c}{24}(N-N^{-1}).
\end{equation}
In the context of symmetric product orbifold (the $\Norb$ in below definition of symmetric product orbifold is in general different from the $N$ as degree of branched cover; see footnote~\ref{footnote-N-difference}.)
\begin{equation}
    \mathcal{C}^{\otimes \Norb}/S_\Norb,
\end{equation}
the (connected) gauge-invariant twist operator correlator is a partition function with prescribed ramification data and admits a representation as summing over the gauge-dependent twist operators $\zm$ over Hurwitz space (the moduli space of branched covers; see Definition~\ref{def-hurwitz-space})~\cite{Pakman:2009zz,Dei:2019iym}:
\begin{equation}
    \mathcal{Z}_{\mathbf{r}}(\bm{z}|\bm{\lambda}) = \expval{\prod_i \lambda_i(z_i)} = \sum_g \mathcal{N}_{g,\Norb}\prn{\vb*{\lambda}} \sum_{\substack{\phi_\md \in \Hurwitz \\ \text{br}\prn{\phi_\md}=\bpts }} \zm\prn{\bpts|\perms}.
\end{equation}
Here ramification data is a pair $\mathbf{r} = (\bm{\lambda},\bm{z}) \in P^M_\Norb \times \mathbb{C}^M$, with $P_\Norb$ being the set of integer partitions of $\Norb$. $\Hurwitz$ is Hurwitz space at genus $g$ with ramification profile $\vb*{\lambda}$ and $\text{br}\prn{\phi}$ denotes the branch locus of a branched cover $\phi$. \footnote{As argument in $\Hurwitz$, $\vb*{\lambda}$ should be viewed as integer partitions of $N$, the degree of branched cover $\phi_\md$, instead of $\Norb$, with $g+N$ fixed by total ramification orders via Riemann-Hurwitz formula. The degree of branched cover $N$ is called the number of “active colors'' among all the $\Norb$ color indices in~\cite{Pakman:2009zz}.\label{footnote-N-difference}} $\mathcal{N}_{g,\Norb}\prn{\vb*{\lambda}}$ is a combinatorial normalization constant whose precise form can be found in, e.g.,~\cite{Dei:2019iym}, and in the large $\Norb$ limit the gauge-invariant twist operator has a genus expansion~\cite{Lunin:2000yv,Pakman:2009zz,Dei:2019iym}:\footnote{This is derived in the case where each partition/cycle structure $\lambda_i$ has only one non-trivial cycle.}
\begin{equation}
    \lim_{\Norb \to \infty}\mathcal{N}_{g,\Norb}\prn{\vb*{\lambda}} \sim \Norb^{1-g-\frac{M}{2}}.
\end{equation}
The genus expansion of symmetric product orbifold at large $\Norb$ limit has motivated studies on its relation with gauge theory/string theory/holography. Despite the rich literature on the quantity, some aspects of the general structure of twist operator correlator, as will be explained below, are not fully understood. One goal of this paper is to clarify the general structures of twist operator correlators associated with generic branched covers of genus zero and one.

A method of computing the twist operator correlator directly from its path integral definition is developed in~\cite{Lunin:2000yv}, where one performs the path integral on the covering surface $\Sigma$ and takes into account the Weyl anomaly due to the Weyl transformation induced by the covering map $\phi_\md$. It is clear from the path integral perspective that for branched cover of genus zero and one, twist operator correlator has the following general structure:
\begin{align}
\label{eq-twist-operator-general-structure}
\zm = 
    \begin{cases}
    \abs{\wm}^{2c}  & g=0 \\
    \abs{\wm}^{2c} \pf(\tau_\md,\Bar{\tau}_\md)  & g=1
    \end{cases}
\end{align}
where in general we use subscript in $\md$ to denote dependence on monodromy data as $F_\md = F_\md(\bm{z}|\bm{\sigma})$, $\wm$ is the holomorphic part of Weyl anomaly factor and $\pf(\tau,\Bar{\tau})$ is the torus partition function of the CFT $\mathcal{C}$ with period $\tau$. The Weyl anomaly factor is determined from branched cover by Liouville action
\begin{align}
    &\log \wm + \log \overline{\mathcal W}_\md= \frac{1}{48 \pi}  S^{\text{reg}}_L[\Phi], \ \ \Phi = \log \phi_\md^\prime(w) + \log \Bar{\phi}_\md^\prime(\Bar{w}) \nonumber\\
    &S_L[\Phi] = \int d^2w \sqrt{\hat{g}} \prn{ \hat{R} \Phi + \frac{1}{2} \hat{g}^{\mu \nu} \partial_\mu \Phi \partial_\nu \Phi}, \ \ \hat{d}s^2 = dw d\Bar{w}
\end{align}
where $w$ is the coordinate on covering space, and we refer to~\cite{Lunin:2000yv} for technical details such as regularization of Liouville action. While the path integral method is conceptually clean and makes clear the general structure of twist operator correlator, it has some disadvantages: i) The $\bpts$-dependence of $\wm$ and in turn $\zm$ appears rather indirectly; the branch locus $\bpts$ come in as coefficients in $\phi_\md$, a rational function for $g=0$ and elliptic function for $g=1$, and one obtains $\wm$ as a function of $\bpts$ upon substituting $\phi_\md$ into the regularized Liouville action $S^{\text{reg}}_L[\Phi]$; ii) the calculation of Liouville action requires careful and somewhat tedious regularization procedure, even in the simple cases of two- and three-point function. One is therefore naturally led to ask if there is a direct characterization of the regularized part of $\wm$ and in turn $\zm$ as a function of $\bpts$, without having to go through the indirect and involved Liouville action calculation.

Indeed, an alternative method exists and is known as the stress-tensor method of~\cite{Dixon:1986qv}. The method can be understood as first using conformal Ward identity to derive following differential equation for twist operator correlator
\begin{align}
\label{eq-stress-tensor-method-diff-eq}
    \partial_{z_i} \log \zm = \Res_{z=z_i} \expval{T(z)}_\md, 
\end{align}
where
\begin{equation}
    \expval{\cdot}_\md \coloneqq \frac{ \expval{ \prn{\cdot} \prod_i \sigma_i (z_i)}}{\expval{\prod_i \sigma_i (z_i)}} = \frac{ \int\limits_{\varphi_I \prn{\xi_i \circ z}  = \varphi_{\sigma_i(I)}(z)}  \brk{D\varphi} \prn{\cdot} e^{-\sum_I S[\varphi_I]}}{\int\limits_{\varphi_I \prn{\xi_i \circ z}  = \varphi_{\sigma_i(I)}(z)} \brk{D\varphi} e^{-\sum_I S[\varphi_I]}}
\end{equation}
and 
\begin{equation}
    T(z) = \sum_I T_I(z)
\end{equation}
is the total stress tensor of $N$ copies of CFT $\mathcal{C}$; one then proceeds by finding $\expval{T(z)}_\md$ and solving the differential equation. In standard orbifold CFT literature~\cite{Dixon:1986qv,Arutyunov:1997gt}, $\expval{T(z)}_\md $ is found by using free field realization and therefore the argument is not universal for generic 2d CFTs. 

However, in light of the universal structure~\eqref{eq-twist-operator-general-structure} clear from the path integral perspective, one is naturally led to ask if the universal structure can be understood directly from the stress-tensor method. To the best of our knowledge, such a  generalized formulation of the stress-tensor method for generic 2d CFTs that makes the universal structure transparent is lacking in literature. Moreover, a general expression of $\expval{T(z)}_\md$ directly in terms of branched cover data and torus partition function of a generic CFT $\mathcal{C}$ for generic branched cover of genus zero and one is also lacking. 

We fill these gaps by generalizing the argument of Calabrese-Cardy~\cite{Calabrese:2004eu} in the context of single interval \renyi entropy, i.e., genus zero branched cover with cyclic monodromy. The insight of~\cite{Calabrese:2004eu} is that $\expval{T(z)}_\md$ may as well be found by first finding the stress tensor one-point function on the covering surface in the uniformizing coordinate and then transforming back to the base coordinate $z$. This argument is universal because it only relies on transformation property of stress-tensor. Employing a generalization of this argument, we have following universal expression of stress tensor one-point function associated with generic branched covers of genus zero and one:
\begin{empheq}[box=\fbox]{align}
    \label{eq-stress-tensor-by-cardy-method}
    g=0: \ \ \expval{T(z)}_\md &= \frac{c}{12} \sum_I \{\psi^I_\md, z \}, \nonumber\\
    g=1: \ \ \expval{T(z)}_\md &= \frac{c}{12} \sum_I \{\psi^I_\md, z \} + 2 \pi i \sum_I  {\prn{\psi^{I}_\md}^\prime}^2 \prn{z} \partial_{\tau_\md} \log \pf \prn{\tau_{\mathbf{m}}} \nonumber\\
    &= \frac{c}{12} \sum_I \{u(\piz), z \} + 2 \pi i \sum_I v^2(\piz)  \partial_{\tau_\md} \log \pf \prn{\tau_{\mathbf{m}}}
\end{empheq}
where $\psi^I_\md$ is the inverse of $\phi_\md$, $u(p)$ is Abel map, $\omega(p) = v(p) dz$ is the differential on covering torus in base coordinate $z \in \cpone$, $\piz$ are pre-images of $z$ under $\phi_\md$, and $\{\cdot, z\}$ denotes Schwarzian derivative with respect to $z$.

This indeed agrees with the general structure~\eqref{eq-twist-operator-general-structure}: the anomalous contributions to stress-tensor one-point function corresponds to the Weyl anomaly terms in~\eqref{eq-twist-operator-general-structure} and the additional thermal energy term at genus one corresponds to the torus partition function term in~\eqref{eq-twist-operator-general-structure}. It is non-trivial to confirm that the logarithmic derivatives of Weyl anomaly and torus partition function with respect to branch locus $\bpts$ indeed agree with residues of the corresponding terms in~\eqref{eq-stress-tensor-by-cardy-method}. The agreement for Weyl anomaly term is essentially explained in~\cite{kokotov2003taufunction} by studying variation of Liouville action with respect to branch locus; we will explain that the agreement of torus partition function term follows from a special case of Rauch variation formula derived in~\cite{kokotov2003taufunction,korotkin2003solution}.

The generalized stress-tensor method for twist operator correlator allows us to recognize its close relation with the tau function on Hurwitz space of Kokotov-Korotkin~\cite{kokotov2003taufunction}. In general, tau function is a central concept in the theory of integrable systems, with canonical examples including the ones associated with KP hierarchy and isomonodromic deformations; see,  e.g.,~\cite{babelon_bernard_talon_2003,harnad_balogh_2021} for introduction. The tau function on Hurwitz space of~\cite{kokotov2003taufunction} is known as essentially a special case of the more general isomonodromic tau function~\cite{Jimbo:1981zz} associated with rank $N$ matrix Fuchsian equations while specializing to quasi-permutation matrix monodromies~\cite{korotkin2003solution}. To give more motivation for the relation, we could have asked the following question: given that the Weyl anomaly contribution to twist operator correlator is universal (i.e., only depending on central charge not on other CFT data) and therefore purely a property of the associated branched cover, is it captured by some canonical algebro-geometric object where one associates a branched cover $\phi_\md$ with a function $F_\md(\bpts|\perms)$ of its monodromy data? The tau function on Hurwitz space of~\cite{kokotov2003taufunction} indeed provides such an object and can be thought of as being defined on a cross-section of Hurwitz space with fixed monodromies $\perms$ while varying branch locus $\bpts$. The tau function on Hurwitz space is defined as 
\begin{equation}
    \partial_{z_i} \log \uptau_\md \coloneqq \frac{1}{12} \Res_{z=z_i} S_\md(z)
\end{equation}
where $S_\md(z)$ is the sum of Bergman projective connection of $\phi_\md$ at pre-images $\piz$ evaluated in base coordinate $z$.\footnote{ While this definition apparently differs from the original one in~\cite{kokotov2003taufunction}, its equivalence will be explained. Also we choose a different normalization for convenience to compare with CFT.} While deferring the precise definition of Bergman projective connection to the main text, here we highlight the structural similarity of above definition for tau function with the defining equation~\eqref{eq-stress-tensor-method-diff-eq} of twist operator correlator using stress-tensor method and admit the fact (see~\eqref{eq-sum-of-bergman-projective-connection}) that the Bergman projective connection $S_\md(z)$ indeed takes the same form as $\expval{T(z)}_\md$ in~\eqref{eq-stress-tensor-by-cardy-method}, except that in the genus one case it has a “thermal energy” term generated by its own “partition function” $\eta^{-1}(\tau)$, which originates from the theta function with odd characteristics in the definition of Bergman kernel in terms of which Bergman projective connection is defined. We thereby have following relations between twist operator correlator $\zm$ and tau function on Hurwitz space $\uptau_\md$ (Theorem \ref{thm-relation-zm-taum}):
\begin{empheq}[box=\fbox]{equation}
    \zm =
    \begin{cases}
    \abs{\uptau_\md}^{2c}  & g=0 \\
    \abs{\uptau_\md}^{2c} \abs{\eta(\tau_\md)}^{2c} \pf\prn{\tau_\md,\Bar{\tau}_\md}  & g=1
    \end{cases}
\end{empheq}
where $c$ is the central charge of CFT $\mathcal{C}$, $\eta(\tau)$ is Dedekind eta function, $\pf(\tau,\Bar{\tau})$ is torus partition function of $\mathcal{C}$, and the period $\tau_\md = \tau_\md(\bpts|\perms)$ of covering torus is viewed as a function of branch locus $\bpts$ (not to be confused with the tau function $\uptau_\md$). Conversely, the tau function on Hurwitz space admits a CFT interpretation as the holomorphic part of twist operator correlator of $c=1$ free boson 
\begin{equation}
    \uptau_\md = \zm^{\text{bos.}} \big|_{\text{holo.}}
\end{equation}
except with a non-modular-invariant partition function corresponding to trace over free boson Fock space. We also comment on the relation between twist operator correlator and isomonodromic tau function (see~\eqref{eq-relation-between-zm-isomonodromy-tau}) via the known relation between tau function on Hurwitz space and isomonodromic tau function~\cite{kokotov2003taufunction,korotkin2003solution}.

\textbf{Note on notations/conventions:} As we use different fonts to denote objects with different meanings, we give a quick reference for notations in Table~\ref{table-notation-reference} to avoid confusion. The subscripts in $\md$ for quantities in Table~\ref{table-notation-reference} are understood as shorthand to denote following dependence on (permutation) monodromy data: $F_\md = F_\md(\bm{z}|\bm{\sigma})$; similarly for matrix monodromy data $\mathfrak{m}$.  Twist operator correlator as defined in Definition~\ref{def-partition-function} is technically divergent and requires UV-cutoff around branched points (e.g., as in Liouville action calculation); it is understood that we always refer to its finite cut-off independent part. We also don't keep track of the overall normalization of twist operator correlator $\zm$ and primarily concern with its dependence on branch locus $\bpts$; in fact, the related tau functions are only defined up to overall constant.

\begin{table}[H]
    \centering
    \begin{tabular}{|c|c|c|}
         \hline
         Notation & Meaning & Definition  \\
         \hline

         \hline
         $\md$ & (permutation) monodromy data & Definition~\ref{def-partition-function} \\
         $\zm$ & twist operator correlator & Definition~\ref{def-partition-function} \\
         $\wm$ &  Weyl anomaly & eq.~\eqref{eq-Weyl-anomaly-definition} \\
         $\uptau_\md$ & tau function on Hurwitz space & eq.~\eqref{eq-tau-function-def} \\
         $\vb*{\tau}_\md$ & period matrix & eq.~\eqref{eq-period-matrix-definition} \\ 
         \hline 

         \hline
         $\mathfrak{m}$ & matrix monodromy data &  below eq.~\eqref{eq-continuation-Fuchsian-matrix} \\
         $\uptau_\mathfrak{m}$ & isomonodromic tau function & eq.~\eqref{eq-ismomono-tau-diff-eq-general} \\
         \hline 

         \hline
         $\pf\prn{\tau,\Bar{\tau}}$ & torus partition function & eq.~\eqref{eq-torus-partition-function-def}\\
         \hline
    \end{tabular}
    \caption{Quick reference for notations. The subscripts in $\md$ for quantities in the table are understood as shorthand to denote following dependence on (permutation) monodromy data: $F_\md = F_\md(\bm{z}|\bm{\sigma})$; similarly for matrix monodromy data $\mathfrak{m}$.}
    \label{table-notation-reference}
\end{table}

The paper is organized as follows. In Section~\ref{sec-RS-background} we review aspects of Riemann surfaces and branched covers relevant for our purpose. In particular, we review the Bergman kernel and Bergman projective connection, and comment on their analogy with 2d CFT; we define the tau function on Hurwitz space in a way that makes manifest its relation with twist operator correlator and explain the equivalence of our definition with the original definition. In Section~\ref{sec-twist-operator} we give a brief review of the path integral method for twist operator correlator, explain the standard stress-tensor method for twist operator correlator and derive the universal expression of stress-tensor one-point function~\eqref{eq-stress-tensor-by-cardy-method} that generalizes the conventional stress-tensor method to generic 2d CFTs, and comment on, among other things, the consistency between path integral and stress-tensor method. We give the precise relation between twist operator correlator and tau function on Hurwitz space in Theorem~\ref{thm-relation-zm-taum} and in turn the relation with isomonodromic tau function in~\eqref{eq-relation-between-zm-isomonodromy-tau}. In Section~\ref{sec-discussion}, we comment on the relation of our results with existing literature and mention some remaining questions/future directions.

\section{Riemann surfaces, branched covers and tau function on Hurwitz space}
\label{sec-RS-background}
In this section we review aspects of Riemann surfaces and branched covers relevant for our purpose and introduce the tau function on Hurwitz space of Kokotov-Korotkin~\cite{kokotov2003taufunction} in a way that makes manifest the analogy with twist operator correlator.
\subsection{Branched covers of Riemann surfaces}
The monodromy data $\md$ in Definition~\ref{def-partition-function} is naturally identified as the monodromy of a branched cover 
\begin{equation}
    \phi_\md: \Sigma \to \mathbb{CP}^1,
\end{equation}
which is a meromorphic function on $\Sigma$. The monodromies $\perms$ of a branched cover $\phi_\md$ are required to satisfy
\begin{itemize}
    \item The monodromy group generated by $\bm{\sigma}$ is a transitive subgroup of $S_N$ \\
    \item Compositions of all permutations in $\bm{\sigma}$ equal identity: $\sigma_1 \cdots \sigma_M = \text{id}$.
\end{itemize}
The first condition imposes w.l.o.g that the covering surface $\Sigma$ is connected, as otherwise the partition function $\mathcal{Z}_{\bf{m}}$ would factorize and one can study each connected component individually. The second condition is necessary as the monodromy data $\bf{m}$ gives a representation of $\pi_1 \prn{\mathbb{CP}^1 \setminus \bm{z}}$.

The genus of a monodromy data $\mathbf{m}$, i.e., genus of covering surface $\Sigma$, is given by Riemann-Hurwitz formula
\begin{equation}
    g= \sum_{\gamma \in \bm{\sigma}} \frac{l-1}{2} -N +1
\end{equation}
where the sum is over all cycles $\gamma$ in cycle decompositions of all permutations in $\bm{\sigma}$ and $l = |\gamma|$ denotes the length of a cycle.

As our notation suggests, we think of the branched cover $\phi_\md$ as parametrized by its monodromy data $\md$; this is justified by Riemann's existence theorem, which essentially states the bijection between the set of inequivalent branched covers/meromorphic functions on $\Sigma$ and the set of monodromy data. Indeed, the twist operator correlator in Definition~\ref{def-partition-function} would not be well-defined if one could find two inequivalent branched covers with, for example, different Liouville actions but same monodromy data.
\begin{theorem}[Riemann's existence theorem\footnote{Apparently related but different statements go by the name Riemann's existence theorem. The exact statement in the theorem can be found in, e.g., Theorem 1.8.14 of~\cite{Lando2003GraphsOS}.}]
Given a monodromy data $\md = (\perms,\bpts)$, there exists a compact Riemann surface $\Sigma$ and a meromorphic function $\phi_\md: \Sigma \to \mathbb{CP}^1$ such that $\bpts$ are branch locus of $\phi_\md$ and $\perms$ are the corresponding permutation monodromies. The branched cover $\phi_\md$ is unique up to isomorphism.
\end{theorem}

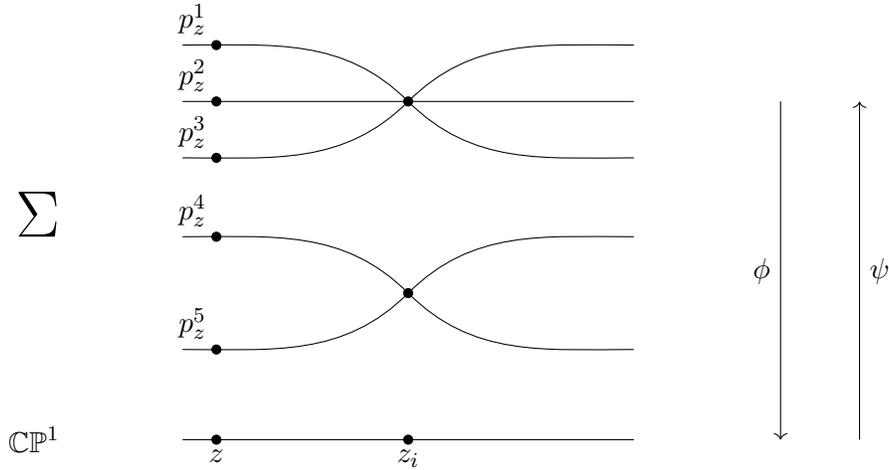
\begin{figure}[h]
    \centering
    \begin{tikzpicture}[scale=1.5]
        \draw [fill] (0,0) circle [radius=0.04];
        \draw (-2,.5) to [out=0,in=135] (0,0);
        \draw (0,0) to [out=-45,in=180] (2,-.5);
        \draw (-2,0) to (2,0);
        \draw (-2,-.5) to [out=0,in=-135] (0,0);
        \draw (0,0) to [out=45,in=180] (2,.5);
        \draw [fill] (-1.7,.5) circle [radius=0.04];
        \draw [fill] (-1.7,0) circle [radius=0.04];
        \draw [fill] (-1.7,-.5) circle [radius=0.04];
        \node [above left] at (-1.7,.5) {$p^1_z$};
        \node [above left] at (-1.7,0) {$p^2_z$};
        \node [above left] at (-1.7,-.5) {$p^3_z$};

        \node [above left] at (-1.7,-1.2) {$p^4_z$};
        \node [above left] at (-1.7,-2.2) {$p^5_z$};
        \draw [fill] (-1.7,-1.2) circle [radius=0.04];
        \draw [fill] (-1.7,-2.2) circle [radius=0.04];
        \draw [fill] (0,-1.7) circle [radius=0.04];
        \draw (-2,-1.2) to [out=0,in=135] (0,-1.7);
        \draw (0,-1.7) to [out=-45,in=180] (2,-2.2);
        \draw (-2,-2.2) to [out=0,in=-135] (0,-1.7);
        \draw (0,-1.7) to [out=45,in=180] (2,-1.2);

        \draw (-2,-3) -- (2,-3);
        \draw [fill] (-1.7,-3) circle [radius=0.04];
        \draw [fill] (0,-3) circle [radius=0.04];
        \node [below] at (-1.7,-3) {$z$};
        \node [below] at (0,-3) {$z_i$};

        \node [left] at (-3,-1) {\Huge$\Sigma$};
        \node [left] at (-3,-3){$\cpone$};

        \draw [->] (3.3,0) -- (3.3,-3);
        \node [left] at (3.3,-1.5) {$\phi$};
        \draw [<-] (4,0) -- (4,-3);
        \node [right] at (4,-1.5) {$\psi$};
    \end{tikzpicture}
    \caption{Visualization of the local structure of a degree $N=5$ branched cover near a branched point $z_i$ with monodromy $\sigma_i = (123)(45)$. The branched cover $\phi$ is a projection and its inverse $\psi$ a lift.}
    \label{fig-branched-cover-local}
\end{figure}

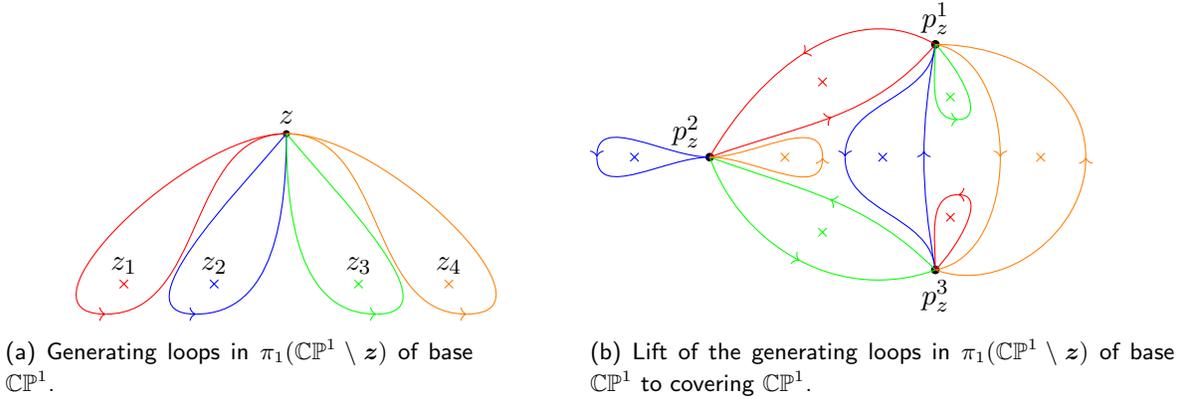
\begin{figure}[h]
    \centering
    \begin{subfigure}[t]{0.4\textwidth}
    \begin{tikzpicture}[scale=.8]
    \draw [fill] (0,0) circle [radius=0.05];
    \node [above] at (0,0) {$z$};
    \draw [->,red] (0,0) to [out=180, in=180] (-3,-3);
    \draw [red] (-3,-3) to [out=0,in=180] (0,0);
    \draw [->,blue] (0,0) to [out=-130, in=180] (-1.5,-3);
    \draw [blue] (-1.5,-3) to [out=0,in=-90] (0,0);
    \draw [->,green] (0,0) to [out=-90, in=180] (1.5,-3);
    \draw [green] (1.5,-3) to [out=0,in=-50] (0,0);
    \draw [->,orange] (0,0) to [out=0, in=180] (3,-3);
    \draw [orange] (3,-3) to [out=0,in=0] (0,0);
    \draw (-2.7,-2.5) node[cross,red]{};
    \node [above] at (-2.7,-2.5) {$z_1$};
    \draw (-1.2,-2.5) node[cross,blue]{};
    \node [above] at (-1.2,-2.5) {$z_2$};
    \draw (1.2,-2.5) node[cross,green]{};
    \node [above] at (1.2,-2.5) {$z_3$};
    \draw (2.7,-2.5) node[cross,orange]{};
    \node [above] at (2.7,-2.5) {$z_4$};
     \end{tikzpicture}
    \caption{Generating loops in $\pi_1(\cpone \setminus \bpts)$ of base $\cpone$.}
    \end{subfigure}
    \hfill
    \begin{subfigure}[t]{0.5\textwidth}
    \begin{tikzpicture}
    \coordinate (p1) at (0,0);
    \coordinate (p2) at (-3,-1.5);
    \coordinate (p3) at (0,-3);
    
    \draw [fill] (p1) circle [radius=0.05];
    \draw [fill] (p2) circle [radius=0.05];
    \draw [fill] (p3) circle [radius=0.05];
    \draw [->-,red] (p1) to [out=150,in=60] (p2);
    \draw [->-,red] (p2) to [out=20,in=-130] (p1);
    \draw [-<-,blue] (p1) to [out=-100,in=100] (p3);
    \draw [->,blue] (p1) to [out=-95,in=90] (-1.2,-1.5);
    \draw [blue] (-1.2,-1.5) to [out=-90,in=95] (p3);
    \draw [->-,green] (p2) to [out=-70,in=-160] (p3);
    \draw [->-,green] (p3) to [out=135,in=-20] (p2);
    \draw [->-,orange] (p1) to [out=-10,in=10] (p3);
    \draw [->,orange] (p3) to [out=-20,in=-90] (2,-1.5);
    \draw [orange] (2,-1.5) to [out=90,in=0] (p1);
    
    \draw [->,red] (p3) to [out=45,in=0] (.3,-2);
    \draw [red] (.3,-2) to [out=180,in=90] (p3);
    \draw [->,green] (p1) to [out=-90,in=180] (.3,-1);
    \draw [green] (.3,-1) to [out=0,in=-45] (p1);
    \draw [->,blue] (p2) to [out=180,in=90] (-4.5,-1.5);
    \draw [blue] (-4.5,-1.5) to [out=-90,in=180] (p2);
    \draw [orange] (p2) to [out=0,in=90] (-1.5,-1.5);
    \draw [<-,orange] (-1.5,-1.5) to [out=-90,in=0] (p2);
    
    \node [above] at (p1) {$p^1_z$};
    \node [below] at (p3) {$p^3_z$};
    \node [above left] at (p2) {$p^2_z$};
    
    \draw (.2,-2.3) node[cross,red]{};
    \draw (-1.5,-0.5) node[cross,red]{};
    \draw (-.7,-1.5) node[cross,blue]{};
    \draw (-4,-1.5) node[cross,blue]{};
    \draw (.2,.-.7) node[cross,green]{};
    \draw (-1.5,-2.5) node[cross,green]{};
    \draw (-2,-1.5) node[cross,orange]{};
    \draw (1.4,-1.5) node[cross,orange]{};
    \end{tikzpicture}
    \caption{Lift of the generating loops in $\pi_1(\cpone \setminus \bpts)$ of base $\cpone$ to covering $\cpone$.}
    \end{subfigure}
    \caption{Visualization of the global structure of a genus zero $M=4, N=3$ branched cover with monodromies $\sigma_1 = (12), \sigma_2 = (13), \sigma_3 = (23), \sigma_4 = (13)$.}
    \label{fig-branched-cover-global}
\end{figure}

We will focus on branched covers of genus zero and one. For genus zero, the meromorphic function $\phi_\md$ is a rational function, and the inverses $\psimdi$ are solutions of a polynomial equation. For genus one, the meromorphic function $\phi_\md$ is an elliptic function, and the inverses are given by Abel map
\begin{equation}
    \psimdi(z) = u(\piz) = \int^{\piz} \omega(p)
\end{equation}
where $\omega(p) = v(p) dz$ is the differential on covering torus in base coordinate $z \in \cpone$, $\piz$ is the $I^{\text{th}}$ pre-image of $z$ on covering torus, and 
\begin{equation}
    \prn{\psimdi}^\prime(z) = v(\piz).
\end{equation}
We will give explicit examples of branched covers with cyclic monodromy and their associated twist operator correlators in Sec.~\ref{sec-twist-operator}; some useful visualizations of branched covers are given in Figure~\ref{fig-branched-cover-local},~\ref{fig-branched-cover-global}.

We will also need the concept of Hurwitz space, the moduli space of branched covers; more details can be found in, e.g.,~\cite{Cavalieri2016RiemannSA}. 
\begin{definition}[Hurwitz space] 
\label{def-hurwitz-space}
Let $\vb*{\lambda} \in P^M_N$ be a set of $M$ integer partitions of $N$. The Hurwitz space $\hurwitz$ is the set of isomorphism classes of genus $g$ degree $N$ connected branched covers
\begin{equation}
    \phi: \Sigma_g \to \cpone
\end{equation}
with $\vb*{\lambda}$ being the ramification profiles at branch locus of $\phi$. As the degree of branched cover is already determined from genus and ramification profiles via Riemann-Hurwitz formula, we simply denote Hurwitz space by $\Hurwitz$.
\end{definition}
The branch locus $\bpts$ of $\phi$ are not fixed in the definition of Hurwitz space $\Hurwitz$; the branch point map, $\text{br}$, is used in math literature to denote the branch locus $\bpts$ of a particular branched cover $\phi$ in Hurwitz space, i.e.,
\begin{equation}
    \text{br}\prn{\phi} = \bpts.
\end{equation}
The Hurwitz space $\Hurwitz$ may be viewed as being parametrized by two set of coordinates: the branch locus $\bpts$ and monodromies $\perms$ (with cycle structures $\vb*{\lambda}$).

\subsection{Bergman kernel and Bergman projective connection}
Here we review the Bergman kernel (also known as fundamental second kind differential, fundamental normalized bidifferential, etc.) in terms of which the Bergman projective connection is defined, and comment on their analogy with free boson 2d CFT. A classic reference on the material is~\cite{Fay1973ThetaFO}; we largely follow the convention in~\cite{Eynard2018LecturesNO}. The analogy with free boson is also observed in the context of hyperelliptic surfaces (cyclic degree two branched covers) in~\cite{Zamolodchikov:1987ae}; see also~\cite{Gavrylenko:2015cea}.
\subsubsection{Bergman kernel}
Given a compact Riemann surface $\Sigma$, the Bergman kernel $B(p,q)$ is the unique meromorphic symmetric (1,1) form on $\Sigma \times \Sigma$ that has a normalized double pole as $p \to q$ and no other poles, i.e., in any local coordinate $x(p)$ it behaves as
\begin{equation}
    B(p,q) = \frac{dx(p) dx(q)}{(x(p)-x(q))^2} + \text{analytic at $q$},
\end{equation}
and satisfies
\begin{align}
    &B(p,q) = B(q,p) \\
    &\oint_{q \in \mathcal{A_\alpha}} B(p,q) = 0 \\
    &\oint_{q \in \mathcal{B_\alpha}} B(p,q) = 2 \pi i \omega_\alpha(p)
\end{align}
where $\mathcal{A_\alpha},\mathcal{B_\alpha}$ are usual homology cycles and $\omega_\alpha(p)$ basis of differentials.
For $g=0$, the Bergman kernel is given by
\begin{equation}
    B(z,z^\prime) = \frac{dz dz^\prime}{(z-z^\prime)^2}, \ \ z,z^\prime \in \mathbb{CP}^1.
\end{equation}
For $g \geq 1$, the Bergman kernel is given by
\begin{equation}
    B(p,q) = d_p d_q \log \Theta_{\vb{c}}\prn{\vb{u}(p) - \vb{u}(q)|\vb*{\tau}}
\end{equation}
where $\Theta_{\vb{c}}(\vb{u}|\vb*{\tau})$ is Riemann theta function with a half-integer odd characteristics $\vb{c}$, $\vb*{\tau}$ is the period matrix of $\Sigma$, $\vb{u}(p)$ is Abel map.\footnote{See Appendix~\ref{sec-theta-function} for review of definitions and conventions of theta functions. The period matrix is defined by \begin{equation}
\label{eq-period-matrix-definition}
    \prn{\vb*{\tau}}_{\alpha \beta} = \oint_{\mathcal{B}_{\alpha}} \omega_\beta.
\end{equation}} For our purpose we focus on the $g=1$ case, where the Abel map gives a uniformizing coordinate on torus $T_\tau = \mathbb{C}/(\mathbb{Z} + \tau \mathbb{Z})$, and the only odd characteristics is $c = \frac{1}{2} + \frac{\tau}{2}$. For $g=1$ the Bergman kernel can be explicitly written as
\begin{align}
    B(u,u^\prime) &= \partial_u \partial_{u^\prime} \log \theta_1 (u - u^\prime | \tau) du du^\prime, \ \ u,u^\prime \in T_\tau \nonumber\\
    &= \prn{\wp(u-u^\prime|\tau) - \frac{1}{3} \frac{\theta^{'''}_1(0|\tau)}{\theta^{'}_1(0|\tau)}} du du^\prime \nonumber\\
    &= \prn{\wp(u-u^\prime|\tau) - \frac{4 \pi i}{3} \partial_\tau \log \theta^{\prime}_1(0|\tau)} du du^\prime
\end{align}
where $\wp(u|\tau)$ is Weierstrass elliptic function
\begin{equation}
    \wp(u|\tau) = \frac{1}{u^2} + \sum_{\lambda \in \Lambda \setminus \{0\}} \prn{\frac{1}{(u-\lambda)^2} - \frac{1}{\lambda^2}}, \quad \Lambda = \mathbb{Z} + \tau \mathbb{Z},
\end{equation}
second equality follows from identity
\begin{equation}
    -\partial^2_u \log \theta_1(u|\tau) = \wp(u|\tau) - \frac{1}{3}\frac{\theta^{'''}_1(0|\tau)}{\theta^{'}_1(0|\tau)},
\end{equation}
and third equality from heat equation
\begin{equation}
    \theta^{''}_1(u|\tau) = 4 \pi i \partial_\tau \theta_1(u|\tau).
\end{equation}
\begin{remark}[Relation with free boson]
The weight (1,1) Bergman kernel at genus zero and one is analogous to the two point function of the weight $h=1$ current operator $j(z) = i \partial X(z)$ in (uncompactified) free boson:
\begin{align}
    &\expval{j(z) j(z^\prime)}_{\cpone} = \frac{1}{(z-z^\prime)^2}\nonumber\\
    &\expval{j(u) j(u^\prime)}_{T_\tau} = \wp(u-u^\prime|\tau) - \frac{4 \pi i}{3} \partial_\tau \log \theta^{\prime}_1(0|\tau) - \frac{\pi}{\Im\tau}
\end{align}
except without the zero mode contribution for genus one two point function.
\end{remark}
\subsubsection{Bergman projective connection for meromorphic function}
Given a meromorphic function $\phi$ on $\Sigma$, the Bergman kernel can be used to define the Bergman projective connection of $\phi$ at a point $p$ on $\Sigma$:
\begin{equation}
    S_\phi(p) \coloneqq 6 \lim_{p^\prime \to p} \prn{ \frac{B(p,p^\prime)}{d\phi(p) d\phi(p^\prime)} - \frac{1}{\prn{\phi(p) - \phi(p^\prime)}^2} }.
\end{equation}
For our purpose, the meromorphic function $\phi$ will be the branched cover
\begin{equation}
    \phi_\md: \Sigma \to \mathbb{CP}^1.
\end{equation}
Viewing the projective connection as a function of base coordinate $z \in \cpone$, we write
\begin{equation}
    S^I_\md(z) \coloneqq S_{\phi_\md}(p^I_z) = 6\lim_{z \to z^\prime} \prn{\frac{B\prn{p^I_z, p^I_{z^\prime}}}{dz dz^\prime} - \frac{1}{(z-z^\prime)^2}}
\end{equation}
where $p^I_z = \psi^I_\md(z) \coloneqq \prn{\phi^I_\md}^{-1}(z)$ is the lift of the point $z$ to the $\ith$ sheet of covering surface $\Sigma$. In general under a change of coordinate $z \mapsto w(z)$, a projective connection transforms as 
\begin{equation}
\label{eq-transformation-propety-projective-connection}
    S(w) = \prn{\frac{dw}{dz}}^{-2} \brkbig{S(z) + \{w,z\}}
\end{equation}
where $\{f,z\}$ is Schwarzian derivative
\begin{equation}
    \{f,z\} = \frac{f^{'''}(z)}{f^{'}(z)} - \frac{3}{2} \prn{\frac{f^{''}(z)}{f^{'}(z)}}^2.
\end{equation}
For a genus zero branched cover $\phi_\md$, the associated Bergman projective connection is given by
\begin{align}
    S^I_\md(z) &= 6 \lim_{z^\prime \to z} \prn{ \frac{\prn{\psimdi}^\prime(z) \prn{\psimdi}^\prime(z^\prime)}{\prn{\psimdi(z) - \psimdi(z^\prime)}^2} - \frac{1}{(z-z^\prime)^2}} \nonumber\\
    &= \{\psimdi,z\}.
\end{align}
For a genus one branched cover $\phi_\md$, the associated Bergman projective connection is given by
\begin{align}
    S^I_\md(z) &=6 \lim_{z^\prime \to z} \cbrk{ \brk{\wp\prnbig{\psimdi(z) - \psimdi(z^\prime)\big|\tau} - \frac{4 \pi i}{3} \partial_{\tau_\md} \log \theta^{\prime}_1(0|\tau_\md)} \prn{\psimdi}^\prime(z)\prn{\psimdi}^\prime(z^\prime) - \frac{1}{(z-z^\prime)^2}} \nonumber\\
    &= \{\psimdi,z\} - 8 \pi i {\prn{\psimdi}^\prime}^2(z) \partial_{\tau_\md} \log \theta^{\prime}_1(0|\tau_\md) \nonumber\\
    &= \{u\prn{\piz},z\} - 8 \pi i v^2(\piz) \partial_{\tau_\md} \log \theta^{\prime}_1(0|\tau_\md).
\end{align}
For convenience, we introduce following notation 
\begin{equation}
    S_\md(z) \coloneqq \sum_I S^I_\md(z)
\end{equation}
and summarize
\begin{empheq}[box=\fbox]{align}
    \label{eq-sum-of-bergman-projective-connection}
    g=0: \ \ S_\md(z) &= \sum_I \{\psi^I_\md, z \}, \nonumber\\
    g=1: \ \ S_\md(z) &= \sum_I \{\psi^I_\md, z \} - 8 \pi i \sum_I  {\prn{\psi^{I}_\md}^\prime}^2 \prn{z} \partial_{\tau_\md} \log \theta^{\prime}_1(0|\tau_\md) \nonumber\\
    &=  \sum_I \{u(\piz), z \} - 8 \pi i \sum_I v^2(\piz)  \partial_{\tau_\md} \log \theta^{\prime}_1(0|\tau_\md).
\end{empheq}

\begin{remark}[Relation with free boson]
The Bergman projective connection $S^I_\md(z)$ associated with a branched cover defined via regularizing diagonal part of Bergman kernel has close analogy with the one-point function of $\ith$ copy of normal-ordered stress-tensor of free boson under monodromy conditions of branched cover. In light of previous remark, the Bergman kernel may be identified as (up to the difference in zero mode term at genus one)
\begin{equation}
    \frac{B\prn{p^I_z, p^I_{z^\prime}}}{dz dz^\prime} = \expval{j_I(z) j_I(z^\prime)}_\md,
\end{equation}
and as the normal-ordered free boson stress-tensor is given by
\begin{equation}
    T^{\text{bos.}}(z) = \lim_{z^\prime \to z} j(z) j(z^\prime) - \frac{1}{(z-z^\prime)^2},
\end{equation}
the Bergman projective connection can therefore be identified, up to overall constant, as 
\begin{equation}
     S^I_\md(z) = \expval{T^{\text{bos.}}_I(z) }_\md
\end{equation}
This is indeed in line with the usual stress-tensor method using free field realization in standard orbifold CFT literature~\cite{Dixon:1986qv,Arutyunov:1997gt}. It will be clear that similar analogy in fact holds for generic 2d CFTs.
\end{remark}

\subsection{Tau function on Hurwitz space}
The tau function on Hurwitz space of~\cite{kokotov2003taufunction} can be defined in terms of the Bergman projective connection $S^I_\md(z)$ associated with branched cover $\phi_\md$ as
\begin{equation}
\label{eq-tau-function-def}
    \partial_{z_i} \log \uptau_\md \coloneqq \frac{1}{12} \Res_{z=z_i} S_\md(z)
\end{equation}
where we have chosen a different normalization for convenience to compare with CFT. The original definition in~\cite{kokotov2003taufunction} corresponds to normalization
\begin{equation}
    \partial_{z_i} \log\uptau^\prime_\md \coloneqq -\frac{1}{6} \Res_{z=z_i} S_\md(z).
\end{equation}
While this in appearance differs from the original definition in~\cite{kokotov2003taufunction}
\begin{equation}
    \partial_{z_i} \log \uptau^\prime_\md \coloneqq -\frac{1}{6} \sum_{\cyc \in \sigma_i} \frac{1}{l(l-2)!} \partial^{l-2}_{x_i} S^\cyc_\md(x_i)\big|_{x_i=0}
\end{equation}
where the sum is over cycles $\cyc$ in the cycle decomposition of $\sigma_i$, $l=|\cyc|$ is the length of the cycle $\cyc$, $x_i$ is the local uniformizing coordinate $x_i = (z-z_i)^{\frac{1}{l}}$ and $S^\cyc_\md(x_i)$ is Bergman projective connection in the local uniformizing coordinate $x_i$ valid near the lift of branch point $z_i$ corresponding to cycle $\cyc$, the two definitions agree and the equivalence can be seen as follows. Consider the contribution of $\Res_{z=z_i} S_\md(z)$ associated with sheets glued in $\cyc$
\begin{align}
    \Res_{z=z_i} \sum_{I \in \cyc} S^I_\md(z) &= \sum_{I \in \cyc} \oint_{\mathcal{C}(z_i)} dz S^I_\md (z) \nonumber\\
    &= \oint_{\mathcal{C}(0)} dx_i \frac{dz}{dx_i} \brk{\prn{\frac{dx_i}{dz}}^2 S^\cyc_\md(x_i) - \{x_i,z \} } \nonumber\\
    &= \oint_{\mathcal{C}(0)} dx_i \frac{dx_i}{dz} S^\cyc_\md(x_i) \nonumber\\
    &= \oint_{\mathcal{C}(0)} dx_i \frac{1}{l} \frac{1}{x^{l-1}_i} S^\cyc_\md(x_i) \nonumber\\
    &= \frac{1}{l(l-2)!} \partial^{l-2}_{x_i} S^\cyc_\md(x_i)\big|_{x_i=0}
\end{align}
where the second equality follows from joining the lifts of loop $\mathcal{C}(z_i)$ to $x_i$ coordinate to obtain the loop $\mathcal{C}(0)$ in $x_i$ coordinate and using transformation property~\eqref{eq-transformation-propety-projective-connection} of projective connection, the third equality from that $\{x_i,z \}$ only has double pole in $z_i$ and therefore doesn't contribute. The two definitions therefore agree upon summing over cycles $\cyc$.

The compatibility condition for the definition of tau function
\begin{equation}
\label{eq-compatibility-tau-function}
    \partial_{z_j} \Res_{z=z_i} S_\md(z) = \partial_{z_i} \Res_{z=z_j} S_\md(z)
\end{equation}
can be verified using Rauch variation formula~\cite{kokotov2003taufunction}.

The tau function on Hurwitz space may be viewed as being defined on a cross-section of Hurwitz space with fixed monodromies $\perms$ while varying branch locus $\bpts$.

\subsection{Rauch variation formula}
In general the Rauch variation formula is concerned with the variation of basis of differentials and period matrix with respect to a Beltrami differential and we refer to~\cite{kokotov2003taufunction,korotkin2003solution} and references therein for its most general form. We will need the following special case of the variation formula to verify the consistency between path integral and stress-tensor method for twist operator correlator.
\begin{theorem}[\cite{kokotov2003taufunction,korotkin2003solution}]
\label{thm-variation-periods}
The variation of the period matrix of a $g \geq 1$ Riemann surface $\Sigma$ realized as branched cover
\begin{equation}
    \phi_\md: \Sigma \to \mathbb{CP}^1 \nonumber
\end{equation}
with respect to the change of branch locus $\bpts$ is given by 
\begin{equation}
    \partial_{z_i} \prn{\vb*{\tau}_\md}_{\alpha \beta} = 2 \pi i \Res_{z=z_i} \sum_I v_\alpha\prn{p^I_z} v_\beta\prn{p^I_z} 
\end{equation}
where $\omega_\alpha(p) = v_\alpha(p)dz$ is the basis of differentials in base space coordinate $z \in \cpone$ and $p^I_z$ are pre-images of $z$ under $\phi_\md$.
\end{theorem}

\section{Twist operator correlator and tau function on Hurwitz space}
\label{sec-twist-operator}
In this section we give a brief review of the path integral method of~\cite{Lunin:2000yv} to highlight the general structure of twist operator correlator, generalize the stress-tensor method of~\cite{Dixon:1986qv} to generic 2d CFTs without relying on free field realization for generic branched covers of genus zero and one, comment on the consistency between two methods, and give the precise relation between twist operator correlator and tau function on Hurwitz space of~\cite{kokotov2003taufunction} in Theorem~\ref{thm-relation-zm-taum} and in turn the relation with isomonodromic tau function.
\subsection{Path integral method}
The crucial observation underlying the path integral method of Lunin-Mathur~\cite{Lunin:2000yv} is that the defining path integral for twist operator correlator in Definition~\ref{def-partition-function} can as well be performed in covering space for a single copy of CFT $\mathcal{C}$ where there's no longer non-trivial boundary conditions. This induces a Weyl transformation on covering space metric
\begin{equation}
    dz d\Bar{z} = e^{\Phi} dw d\Bar{w}, \ \ \Phi = \log \phi_\md^\prime(w) + \log \Bar{\phi}_\md^\prime(\Bar{w})
\end{equation}
with 
\begin{align}
    &g=0: \ \ w \in \cpone \nonumber\\
    &g=1: \ \ w \in T_\tau = \mathbb{C}/(\mathbb{Z} + \tau \mathbb{Z}).
\end{align}
The Weyl transformation leads to Weyl anomaly factor
\begin{equation}
    \mathcal{Z}_{\bf{m}} = \mathcal{Z}\brk{e^{\Phi} \hat{g}}  = \abs{\wm}^{2c}  \mathcal{Z}\brk{\hat{g}}
\end{equation}
with the Weyl anomaly factor given by regularized Liouville action
\begin{align}
\label{eq-Weyl-anomaly-definition}
    &\log \wm + \log \overline{\mathcal W}_\md= \frac{1}{48 \pi}  S^{\text{reg}}_L[\Phi], \quad \Phi = \log \phi_\md^\prime(w) + \log \Bar{\phi}_\md^\prime(\Bar{w})\nonumber\\
    &S_L[\Phi] = \int d^2w \sqrt{\hat{g}} \prn{ \hat{R} \Phi + \frac{1}{2} \hat{g}^{\mu \nu} \partial_\mu \Phi \partial_\nu \Phi}, \ \ \hat{d}s^2 = dw d\Bar{w}.
\end{align}
The path integral on the covering space $\mathcal{Z}[\hat{g}]$ is given by 
\begin{equation}
    \mathcal{Z}[\hat{g}]  = 
    \begin{cases}
    1 & g=0 \\
    \pf(\tau_\md,\Bar{\tau}_\md) & g=1
    \end{cases}.
\end{equation}
Here in the genus zero case, $\mathcal{Z}[\hat{g}]$ has trivial dependence on branch locus $\bpts$; as we are only interested in the $\bpts$-dependence of $\zm$, $\mathcal{Z}[\hat{g}]$ can be set to unity. In the genus one case, $\mathcal{Z}[\hat{g}]$ gives path integral on covering torus and contains non-trivial dependence on $\bpts$ via the dependence of the period of the covering torus on monodromy data $\md$, which we emphasize by writing $\tau_\md = \tau_\md(\bpts|\perms)$, and $\pf\prn{\tau}$ is the torus partition function of the CFT $\mathcal{C}$
\begin{equation}
\label{eq-torus-partition-function-def}
\pf\prn{\tau,\Bar{\tau}} = \Tr \prn{q^{L_0 - \frac{c}{24}} \Bar{q}^{\Bar{L}_0 - \frac{c}{24}} }, \ \ q=e^{2 \pi i \tau}, \ \ \Bar{q} = e^{- 2 \pi i \Bar{\tau}}.
\end{equation}
As mentioned in the introduction, the $\bpts$-dependence of $\wm$ in the path integral method appears rather indirectly: the branch locus $\bpts$ come in as coefficients in $\phi_\md$, a rational function for $g=0$ and elliptic function for $g=1$, and one obtains $\wm$ as a function of $\bpts$ upon substituting $\phi_\md$ into the regularized Liouville action $S^{\text{reg}}_L[\Phi]$. While referring to~\cite{Lunin:2000yv} for details on technicalities such as regularization of Liouville action and examples of computations using the method, we highlight the general structure made clear by the path integral approach:
\begin{equation}
\zm = 
    \begin{cases}
    \abs{\wm}^{2c}  & g=0 \\
    \abs{\wm}^{2c} \pf(\tau_\md,\Bar{\tau}_\md)  & g=1
    \end{cases}
\end{equation}
where at genus zero the universal (i.e., only depending on central charge and not on other CFT data) Weyl anomaly contribution is the only non-trivial contribution, and at genus one there is an additional contribution from torus partition function of the CFT $\mathcal{C}$.

\subsection{Stress-tensor method generalized}
The stress-tensor method of~\cite{Dixon:1986qv} can be understood as follows. One starts by considering the stress-tensor one-point function $\expval{T(z)}_\md$ defined by
\begin{equation}
    \expval{\cdot}_\md \coloneqq \frac{ \expval{ \prn{\cdot} \prod_i \sigma_i (z_i)}}{\expval{\prod_i \sigma_i (z_i)}} = \frac{ \int\limits_{\varphi_I \prn{\xi_i \circ z}  = \varphi_{\sigma_i(I)}(z)}  \brk{D\varphi} \prn{\cdot} e^{-\sum_I S[\varphi_I]}}{\int\limits_{\varphi_I \prn{\xi_i \circ z}  = \varphi_{\sigma_i(I)}(z)} \brk{D\varphi} e^{-\sum_I S[\varphi_I]}},
\end{equation}
with 
\begin{equation}
    T(z) = \sum_I T_I(z)
\end{equation}
being the sum of stress-tensor of $N$ copies of the CFT $\mathcal{C}$. Then from conformal Ward identity
\begin{align}
    \expval{T(z) \prod_i \sigma_i (z_i)} =  \sum_i \prn{\frac{h_{\sigma_i}}{(z - z_i)^2} + \frac{1}{z-z_i} \partial_{z_i}} \expval{ \prod_i \sigma_i (z_i)} \nonumber
\end{align}
it follows that
\begin{equation}
\label{eq-stress-tensor-in-ward-identity}
    \expval{T(z)}_\md = \sum_i \frac{h_{\sigma_i}}{(z - z_i)^2} + \frac{\partial_{z_i} \log \mathcal{Z}_{\mathbf{m}} }{z - z_i}.
\end{equation}
This allows one to derive a differential equation directly characterizing $\zm$ as functions of branch locus $\bpts$ from singularities of $\expval{T(z)}_\md$:
\begin{equation}
    \partial_{z_i} \log \zm = \Res_{z=z_i} \expval{T(z)}_\md.
\end{equation}

We note that at this point the discussion is entirely general and in principle the differential equation holds for generic branched covers with arbitrary genus. To proceed one then needs to find the stress-tensor one-point function $\tonept$ to solve the differential equation and now we restrict the discussion to branched covers of genus zero and one. In standard orbifold CFT literature~\cite{Dixon:1986qv,Arutyunov:1997gt}, the stress-tensor one-point function is obtained using free field realization of twist operators and the method relies on free-field-specific properties of stress-tensors. However, since the path integral method holds for generic 2d CFTs and makes clear the universal structure of twist operator correlator, one would expect that a correspondingly universal method should exist for finding the stress-tensor one-point function and makes clear the same universal structure. To achieve this, we employ a generalization of the argument in~\cite{Calabrese:2004eu} in the context of single interval \renyi entropy (genus zero branched cover with cyclic monodromy). The key observation of~\cite{Calabrese:2004eu} is that one can find the stress tensor one-point function by first evaluating it in the uniformizing coordinate and then transforming back to the base coordinate, which only relies on the universal transformation property of stress-tensor. Generalizing this argument to generic genus zero and one branched covers with non-abelian monodromy, we first find the stress-tensor one-point function of a copy of CFT in the uniformizing coordinate $\psimdi(z)$
\begin{equation}
    \expval{T\prn{\psimdi(z)}} = 
    \begin{cases}
    0 & g=0 \\
    2 \pi i \partial_{\tau_\md} \log \pf\prn{\tau_\md} & g=1
    \end{cases}
\end{equation}
where in the genus zero case it vanishes on covering sphere and in the genus one case it has a thermal energy on covering torus; then from transformation property of stress-tensor
\begin{equation}
        T(w) = \prn{\frac{dw}{dz}}^{-2} \brk{T(z) - \frac{c}{12}\{w,z\}},
\end{equation}
we obtain the stress-tensor one-point function of each copy of CFT in base coordinate
\begin{align}
    g=0: \ \ \expval{T_I(z)}_\md &= \frac{c}{12}  \{\psi^I_\md, z \}, \nonumber\\
    g=1: \ \ \expval{T_I(z)}_\md &= \frac{c}{12}  \{\psi^I_\md, z \} + 2 \pi i  {\prn{\psi^{I}_\md}^\prime}^2 \prn{z} \partial_{\tau_\md} \log \pf \prn{\tau_{\mathbf{m}}} \nonumber\\
    &= \frac{c}{12}  \{u(\piz), z \} + 2 \pi i  v^2(\piz)  \partial_{\tau_\md} \log \pf \prn{\tau_{\mathbf{m}}}
\end{align}
where $\psi^I_\md$ is the inverse of $\phi_\md$, $u(p)$ is Abel map, $\omega(p) = v(p) dz$ is the differential on covering torus in base coordinate $z \in \cpone$ and $\piz$ are pre-images of $z$ under $\phi_\md$. The expression~\eqref{eq-stress-tensor-by-cardy-method} is obtained upon summing over copies of CFT. As will be remarked below, this indeed leads to the same universal structure of twist operator correlator as made clear in the path integral method.

\begin{remark}[Twist operator dimension]
The universal twist operator dimension $h_\sigma$ for a generic $\sigma \in S_N$,
\begin{equation}
\label{eq-twist-operator-dimension}
    h_\sigma = \frac{c}{24} \sum_{\cyc \in \sigma} l- l^{-1},
\end{equation}
where the sum is over cycles $\cyc$ in the cycle decomposition of $\sigma$ and $l = \abs{\cyc}$ is the length of a cycle $\cyc$, can be read off from the genus zero case of~\eqref{eq-stress-tensor-in-ward-identity}
\begin{equation}
\label{eq-genus-zero-diff-eq}
    \sum_i \frac{h_{\sigma_i}}{(z - z_i)^2} + \frac{\partial_{z_i} \log \mathcal{Z}_{\mathbf{m}} }{z - z_i} = \frac{c}{12} \sum_I \{\psi^I_\md, z \}
\end{equation}
as follows. For a local coordinate of the form $\psi(z) = (z-z_i)^{\frac{1}{l}} f(z)$, the double pole term in Schwarzian derivative is fixed by local ramification order
\begin{equation}
    \cbrk{(z-z_i)^{\frac{1}{l}} f(z), z} = \frac{(1-l^{-2})/2}{(z-z_i)^2} + \mathcal{O}\prn{(z-z_i)^{-1}}
\end{equation}
therefore in the Schwarzian derivative term of~\eqref{eq-genus-zero-diff-eq}, the double pole contribution from a cycle $\cyc$ is
\begin{equation}
    \Res_{z=z_i} (z-z_i) \sum_{I \in \cyc} \{\psimdi, z \}  = l \cdot \frac{1-l^{-2}}{2} = \frac{l - l^{-1}}{2} 
\end{equation}
where the extra factor of $l$ takes into account the number of sheets glued at a branch point. The total contribution sums over cycles and gives $h_\sigma$.
\end{remark}

\begin{remark}[Consistency between path integral and stress-tensor method]
\label{remark-consistency-condition}
Comparing between the path integral method and the generalized stress-tensor method leads to following consistency conditions
\begin{align}
    g=0: \ \ \partial_{z_i} \log \wm &= \frac{1}{12} \sum_I \Res_{z=z_i}\{\psi^I_\md, z \}, \\
    g=1: \ \ \partial_{z_i} \log \wm &= \frac{1}{12} \Res_{z=z_i}\sum_I \{\psi^I_\md, z \} = \frac{1}{12} \sum_I \Res_{z=z_i} \{u(\piz), z \}, \\
    \partial_{z_i} \log \pf(\tau_\md) &= 2\pi i \prn{\Res_{z=z_i} \sum_I {\prn{\psi^{I}_\md}^\prime}^2 \prn{z}} \partial_{\tau_\md} \log \pf \prn{\tau_{\mathbf{m}}} \nonumber\\
    &= 2 \pi i \prn{\Res_{z=z_i}\sum_I v^2(\piz)} \partial_{\tau_\md} \log \pf \prn{\tau_{\mathbf{m}}} 
\end{align}
While above identifications are conceptually clear: the Weyl anomaly term in the path integral method corresponds to the anomalous contribution to stress-tensor one-point function in stress-tensor method and the torus partition function term to the thermal energy term, it is non-trivial to directly verify the consistency conditions. The consistency condition for the Weyl anomaly term is essentially verified by studying the variation of Liouville action with respect to branch locus $\bpts$ in~\cite{kokotov2003taufunction} to which we refer for details. The consistency condition involving torus partition function is guaranteed by the $g=1$ case of the variation formula in Theorem~\ref{thm-variation-periods}:
\begin{equation}
\label{eq-variation-periods-genus-one}
    \partial_{z_i} \tau_\md = 2 \pi i \Res_{z=z_i}\sum_I v^2(\piz).
\end{equation}
Alternatively, one may interpret that requiring consistency between path integral and stress-tensor method gives a physical derivation of above variation formula. 
\end{remark}

Below we give concrete examples for calculation of twist operator correlators using the generalized stress-tensor method, to demonstrate its consistency with known results in cases associated with branched covers with cyclic monodromy and note its particular simplicity in the genus one case compared with usual derivation in literature~\cite{Dixon:1986qv,Lunin:2000yv}.
\begin{example}[Cyclic monodromy, genus zero \cite{Calabrese:2004eu}; $N^{\text{th}}$ \renyi entropy of single interval]
Consider genus zero $M=2$ branched cover with monodromies $\perms$
\begin{equation}
\sigma_1 = \sigma^{-1}_2 = (1\dots N),
\end{equation}
the covering map and uniformizing map can be written as
\begin{align}
    \phi_\md(w) &= \frac{z_2 w^N - z_1}{w^N -1}, \nonumber\\
    \psimdi(z) &= \prn{\frac{z-z_1}{z-z_2}}^{\frac{1}{N}} e^{\frac{2 \pi i}{N} I}, \ \ I = 1, \cdots, N.
\end{align}
In this case the stress-tensor method gives
\begin{align}
    \partial_{z_1} \log \zm &= \frac{c}{12} \Res_{z=z_1} \sum_I \{\psi^I_\md, z \} \nonumber\\
    &= \frac{N c}{12}  \Res_{z=z_1} \cbrk{\prn{\frac{z-z_1}{z-z_2}}^{\frac{1}{N}},z } \nonumber\\
    &= - \frac{c}{12}\prn{N - N^{-1}} z^{-1}_{12}
\end{align}
with logarithmic derivative with respect to $z_2$ related by permutation; this therefore gives the usual two-point function
\begin{equation}
    \zm = z^{-2h}_{12} \times \antiholo, \ \ h=\frac{c}{24}(N-N^{-1}).
\end{equation}
\end{example}

\begin{example}[Cyclic monodromy, genus one~\cite{Dixon:1986qv,Lunin:2000yv}; second \renyi entropy of two intervals]
Consider genus one $M=4,N=2$ branched cover with monodromies
\begin{equation}
    \sigma_i = (12), \ \ i=1, \dots, 4.
\end{equation}
The differential in base space coordinate is given by
\begin{align}
    \omega(p^{I}_z) &= (-1)^{I} v(z) dz, \ \ I=1,2 \nonumber\\
    v(z) &= \brk{\prod^4_{i=1} \prn{z-z_i} }^{-\frac{1}{2}}, 
\end{align}
and the period of covering torus is given by
\begin{equation}
\label{eq-period-cyclic}
    \tau_\md = i \frac{{}_2F_1\prn{\frac{1}{2},\frac{1}{2} ;1 ; 1-r }}{{}_2F_1\prn{\frac{1}{2},\frac{1}{2} ;1 ; r }}, \ \ r=\frac{z_{12} z_{34}}{z_{14} z_{32}}.
\end{equation}
Using stress-tensor method, the Weyl anomaly term satisfies
\begin{align}
    \partial_{z_i} \log \wm &= \frac{1}{12} \sum_I \Res_{z=z_i} \{u(\piz), z \} \nonumber\\
    &= \frac{1}{12} \cdot 2 \cdot \Res_{z=z_i} \brk{\frac{v^{\prime\prime}(z)}{v(z)} - \frac{3}{2} \prn{\frac{v^\prime(z)}{v(z)}}^2 } \nonumber\\
    &= - \frac{1}{24} \sum_{j \neq i} z^{-1}_{ij},
\end{align}
and therefore
\begin{equation}
    \wm = \prn{\prod_{i < j} z_{ij}}^{-\frac{1}{24}}.
\end{equation}
The full answer for twist operator correlator is 
\begin{equation}
    \zm = \abs{\prod_{i < j} z_{ij}}^{-\frac{c}{12}} \pf(\tau_\md,\Bar{\tau}_\md)
\end{equation}
where the period $\tau_\md$ as function of branch locus $\bpts$ is given in~\eqref{eq-period-cyclic} and we have used the consistency between path integral and stress-tensor method as remarked previously to obtain the partition function term. We note that unlike usual derivation in literature~\cite{Dixon:1986qv,Lunin:2000yv}, we didn't need the elliptic function covering map $\phi_\md$, whose explicit form for special configuration of branch locus $
\bpts$ can be found in ~\cite{Dixon:1986qv,Lunin:2000yv} and generic configuration in~\cite{Eberhardt:2020akk}. 
\end{example}

\subsection{Relation between twist operator correlator and tau function on Hurwitz space}
We hope that at this point the structural similarity between the defining equation of twist operator correlator in stress-tensor method and the definition of tau function on Hurwitz space in terms of Bergman projective connection
\begin{align}
    \partial_{z_i} \log \zm &= \Res_{z=z_i} \expval{T(z)}_\md, \nonumber\\
    \partial_{z_i} \log \uptau_\md &= \frac{1}{12} \Res_{z=z_i} S_\md(z),
\end{align}
and the similar expressions, reproduced below for convenience of comparison, for stress-tensor one-point function $\expval{T(z)}_\md$ 
\begin{align}
    g=0: \ \ \expval{T(z)}_\md &= \frac{c}{12} \sum_I \{\psi^I_\md, z \}, \nonumber\\
    g=1: \ \ \expval{T(z)}_\md &= \frac{c}{12} \sum_I \{\psi^I_\md, z \} + 2 \pi i \sum_I  {\prn{\psi^{I}_\md}^\prime}^2 \prn{z} \partial_{\tau_\md} \log \pf \prn{\tau_{\mathbf{m}}} \nonumber\\
    &= \frac{c}{12} \sum_I \{u(\piz), z \} + 2 \pi i \sum_I v^2(\piz)  \partial_{\tau_\md} \log \pf \prn{\tau_{\mathbf{m}}},
\end{align}
and Bergman projective connection $S_\md(z)$
\begin{align}
    g=0: \ \ S_\md(z) &= \sum_I \{\psi^I_\md, z \}, \nonumber\\
    g=1: \ \ S_\md(z) &= \sum_I \{\psi^I_\md, z \} - 8 \pi i \sum_I  {\prn{\psi^{I}_\md}^\prime}^2 \prn{z} \partial_{\tau_\md} \log \theta^{\prime}_1(0|\tau_\md) \nonumber\\
    &=  \sum_I \{u(\piz), z \} - 8 \pi i \sum_I v^2(\piz)  \partial_{\tau_\md} \log \theta^{\prime}_1(0|\tau_\md),
\end{align}
has made the relation between $\mathcal{Z}_\md$ and $\uptau_\md$ transparent. In particular, the Bergman projective connection of branched cover $\phi_\md$ at $\ith$ pre-image $\piz$ evaluated in base coordinate $z$, $ S^I_\md(z) = S_{\phi_\md}(p^I_z)$, is analogous to the stress-tensor one-point function of $\ith$ copy of the CFT under monodromy conditions $\md$ of branched cover $\phi_\md$, $\expval{T_I(z)}_\md$; it contains an anomalous term responsible for “Weyl anomaly” and in the genus one case a “thermal energy” term generated by its own “partition function” $\theta^{\prime}_1(0|\tau)^{-\frac{1}{3}} \propto \eta^{-1}(\tau)$, which originates from the theta function with odd characteristics in the definition of Bergman kernel.\footnote{Recall Jacobi's identity $\theta^{\prime}_1(0|\tau) = \theta_2(\tau) \theta_3(\tau) \theta_4(\tau) = 2 \eta^3(\tau)$.} We can therefore write the twist operator correlator and tau function on Hurwitz space in a similar way
\begin{align}
\label{eq-zm-taum-common-parametrization}
    g=0: \ \ \zm &= \abs{\wm}^{2c},  \nonumber\\
    \uptau_\md &= \wm, \nonumber\\
    g=1: \ \ \zm &= \abs{\wm}^{2c} \pf\prn{\tau_\md,\Bar{\tau}_\md},  \nonumber\\  \uptau_\md  &= \wm \eta^{-1}(\tau_\md),
\end{align}
where we have used the compatibility conditions in Remark~\ref{remark-consistency-condition} to identify the Weyl anomaly $\wm$ and torus partition function $\pf\prn{\tau_\md,\Bar{\tau}_\md}$; also the genus one variation formula~\eqref{eq-variation-periods-genus-one} again allows one to directly identify the “partition function” term $\theta^{\prime}_1(0|\tau)^{-\frac{1}{3}} \propto \eta^{-1}(\tau)$  in $\uptau_\md$ as explained in~\cite{kokotov2003taufunction}. We then recognize that in the genus zero and one case the tau function on Hurwitz space is essentially the holomorphic part of the twist operator correlator of $c=1$ free boson, 
\begin{equation}
    \uptau_\md = \zm^{\text{bos.}} \big|_{\text{holo.}},
\end{equation}
except with a non-modular-invariant partition function
\begin{equation}
    \pf^\prime_{\text{bos.}}(\tau,\Bar{\tau}) = |\eta(\tau)|^{-2}
\end{equation}
corresponding to trace over free boson Fock space, instead of the modular-invariant one
\begin{equation}
    \pf_{\text{bos.}}(\tau,\Bar{\tau}) = \prn{\Im\tau}^{-\frac{1}{2}} |\eta(\tau)|^{-2}.
\end{equation}
This is indeed also expected from previous remarks on the analogy between Bergman kernel/projective connection and free boson. 

We summarize the direct relation between twist operator correlator and tau function on Hurwitz space in following theorem, which can be immediately inferred  from~\eqref{eq-zm-taum-common-parametrization}.
\begin{tcolorbox}
\begin{theorem}
\label{thm-relation-zm-taum}
Let $\zm$ be the twist operator correlator in Definition~\ref{def-partition-function} and $\uptau_\md$ be the tau function on Hurwitz space defined in~\eqref{eq-tau-function-def}, both associated with a  branched cover $\phi_\md : \Sigma \to \cpone$ with monodromy data $\md$; then for a generic branched cover $\phi_\md$ of genus zero and one, 
\begin{equation}
    \zm =
    \begin{cases}
\abs{\uptau_\md}^{2c}  & g=0 \\
    \abs{\uptau_\md}^{2c} \abs{\eta(\tau_\md)}^{2c} \pf\prn{\tau_\md,\Bar{\tau}_\md}  & g=1
    \end{cases}
\end{equation}
where $c$ is the central charge of the CFT $\mathcal{C}$ in Definition~\ref{def-partition-function}, $\pf(\tau,\Bar{\tau})$ is its torus partition function, $\eta(\tau)$ is Dedekind eta function and $\tau_\md$ is the period of the covering torus.
\end{theorem}
\end{tcolorbox}

\begin{remark}[Compatibility condition]
The relation between Bergman projective connection and stress-tensor one-point function also makes clear the compatibility condition of~\eqref{eq-stress-tensor-method-diff-eq}
\begin{equation}
    \partial_{z_j} \Res_{z=z_i} \expval{T(z)}_\md =  \partial_{z_i} \Res_{z=z_j} \expval{T(z)}_\md
\end{equation}
from the known compatibility condition~\eqref{eq-compatibility-tau-function} for Bergman projective connection $S_\md(z)$, as they essentially coincide in the genus zero case and only differ in the genus one case by terms that trivially satisfy the compatibility condition.
\end{remark}

\begin{remark}[Relation with isomonodromic tau function~\cite{korotkin2003solution}]
The tau function on Hurwitz space is known essentially as special case of the more general isomonodromic tau function associated with rank $N$ matrix Fuchsian equation with $M$ singularities
\begin{equation}
    \partial_z \Psi(z) = A(z) \Psi(z), \ \ A(z) = \sum_i \frac{A_i}{z-z_i}, \ \ A_i \in \text{GL}(N,\mathbb{C}), \ \ \Tr\prn{A_i}=0,
\end{equation}
where the matrix function $\Psi(z) \in \text{GL}(N,\mathbb{C})$ has monodromies
\begin{equation}
\label{eq-continuation-Fuchsian-matrix}
     \Psi(\xi_i \circ z) = \Psi(z) M_i, \ \ M_i \in \text{SL}(N,\mathbb{C})\\
\end{equation}
with monodromy matrices satisfying $M_M \cdots M_1 = \mathbb{I}$ again because the matrix monodromy data  $\mathfrak{m} = \prn{\bm{M},\bm{z}} \in \text{SL}(N,\mathbb{C})^M \times \mathbb{C}^M$ gives a representation of $\pi_1 \prn{\mathbb{CP}^1 \setminus \bm{z}}$.

Isomonodromic deformation of such matrix Fuchsian equation is concerned with changing $A_i \in \bm{A}$ as function of $\bm{z}$ while keeping $\bm{M}$ fixed. The isomonodromic deformation is governed by a set of non-linear PDEs known as Schlesinger equations
\begin{align}
    \partial_{z_j} A_i &= \frac{\comm{A_i}{A_j}}{z_{ij}}, \ \ i \neq j \nonumber\\
    \partial_{z_i} A_i &= - \sum_{j \neq i} \frac{\comm{A_i}{A_j}}{z_{ij}},
\end{align}
and given a solution to Schlesinger equation, the associated isomonodromic tau function is defined in terms of the solution as~\cite{Jimbo:1981zz}
\begin{align}
\label{eq-ismomono-tau-diff-eq-general}
        \partial_{z_i} \log \uptau_{\mathfrak{m}}\prn{\bm{z}|\bm{M}} &\coloneqq \frac{1}{2} \Res_{z=z_i} \Tr\prn{A^2}. 
\end{align}
It is shown in~\cite{korotkin2003solution} that the tau function on Hurwitz space is essentially the isomonodromic tau function while specializing to quasi-permutation monodromy matrices
\begin{equation}
    (M_i)_{IJ} = \pm \delta_{I,\sigma_i(J)}
\end{equation}
where the minus signs arise from branch cuts on covering space and we refer to~\cite{korotkin2003solution} for details, and the two tau functions essentially coincide up to theta function
\begin{equation}
 \uptau_\mathfrak{m} = \uptau_\md \Theta\prn{0|\vb*{\tau}_\md}
\end{equation}
where $\vb*{\tau}_\md$ is the period matrix of covering surface and at genus zero the two tau functions exactly coincide.\footnote{The construction in~\cite{korotkin2003solution} in fact involves sets of additional parameters; the isomonodromic tau function we consider here corresponds to not turning on such parameters.} Therefore, by virtue of the relation in Theorem~\ref{thm-relation-zm-taum} we have following relation between twist operator correlator and isomonodromic tau function
\begin{equation}
\label{eq-relation-between-zm-isomonodromy-tau}
    \zm =
    \begin{cases}
    \abs{\uptau_\mathfrak{m}}^{2c}  & g=0 \\
    \abs{\uptau_\mathfrak{m}}^{2c} \abs{\frac{\eta\prn{\tau_\md}}{\theta_3 \prn{\tau_\md}} }^{2c} \pf\prn{\tau_\md,\Bar{\tau}_\md}  & g=1
    \end{cases}.
\end{equation}
We note that now conversely the isomonodromic tau function at genus zero and one admits a CFT interpretation as the holomorphic part of the associated twist operator correlator of two copies of $c=\frac{1}{2}$ free fermion with the non-modular-invariant torus partition function with (NS,NS) boundary conditions
\begin{equation}
    \pf^{\text{(NS,NS)}}_{\text{ferm.}} \prn{\tau,\Bar{\tau}} = \abs{ \frac{\theta_3(\tau)}{\eta(\tau)}}.
\end{equation}
\end{remark}

\begin{remark}[Integrable system interpretation]
The relation with tau function gives a quite different interpretation of the Weyl anomaly contribution to twist operator correlator: while in the context of twist operator correlator the branch locus are viewed as locations of operator insertions and monodromies as boundary conditions, in the context of isomonodromic deformation the branch locus can be viewed as “times” and the monodromies kept fixed during the deformation as “conserved quantities”.
\end{remark}

\begin{remark}[Gauge-invariant twist operator]
As mentioned in introduction, the gauge-invariant twist operator correlator $\mathcal{Z}_{\mathbf{r}}(\bm{z}|\bm{\lambda})$ in symmetric product orbifold $\mathcal{C}^{\otimes \Norb}/S_\Norb$ admits a representation as summing over the gauge-dependent twist operators $\zm$ over Hurwitz space and has a genus expansion in the large $\Norb$ limit. In light of the relation in Theorem~\ref{thm-relation-zm-taum} between gauge-dependent twist operator and tau function on Hurwitz space, the leading order genus zero and one contribution at large $\Norb$ limit of gauge-invariant twist operator therefore admits a representation as summing over tau functions on Hurwitz space:
\begin{align}
    \mathcal{Z}_{\vb{r}}(\bpts | \bm{\lambda}) &= \mathcal{N}_{0,\Norb}\prn{\vb*{\lambda}} \sum_{\substack{\phi_\md \in \mathcal{H}_{0}\prn{\vb*{\lambda}} \\ \text{br}\prn{\phi_\md}=\bpts }} \abs{\uptau_\md\prn{\bpts | \perms}}^{2c}  \nonumber\\
    &+ \mathcal{N}_{1,\Norb}\prn{\vb*{\lambda}} \sum_{\substack{\phi_\md \in \mathcal{H}_{1}\prn{\vb*{\lambda}} \\ \text{br}\prn{\phi_\md}=\bpts }} \abs{\uptau_\md\prn{\bpts | \perms}}^{2c} \abs{\eta (\tau_\md)}^{2c} \pf\prn{\tau_\md,\Bar{\tau}_\md}  + \cdots
\end{align}
\end{remark}

\section{Discussions}
\label{sec-discussion}
We conclude by commenting on the relation between our results with existing literature and mentioning some open questions we hope to address in the future.

\subsection{Relation with existing literature}
\begin{itemize}
    \item \textit{Twist operator correlator and tau function on Hurwitz space.} The relation between twist operator correlator in 2d CFT and tau function on Hurwitz space is also studied  in~\cite{Gavrylenko:2015cea} in the context of interpreting tau function as twist operator conformal blocks of W-algebra by utilizing free field realization of $c=N-1$ $W_N$ algebra. While similar technical details are discussed such as the analogy between Bergman projective connection and stress-tensor in 2d CFT, our purpose and perspective are quite different: we relate two independently well-defined objects, the universal Weyl anomaly contribution to twist operator correlator and tau function on Hurwitz space, and concern generic 2d CFTs without relying on free-field realization. Also the Liouville action associated with Weyl anomaly contribution to twist operator correlator is studied in the original paper~\cite{kokotov2003taufunction} on tau function on Hurwitz space where it is observed that the Liouville action  solves (the non-trivial part of) the defining differential equation for tau function (i.e., the Weyl anomaly part of consistency conditions in Remark~\ref{remark-consistency-condition}); we clarify the physical origin of Liouville action in their tau function calculation by pointing out the relation between tau function and twist operator correlator.
    
    \item \textit{Twist operator correlator and isomonodromic tau function.} The relation between twist operator correlator and isomonodromic tau function dates back to the holonomic quantum fields of~\cite{Sato:1979kg} in the early studies of isomonodromic tau function, where solution of matrix Riemann-Hilbert problem with generic matrix monodromy data is constructed using twist operators in free fermion and the isomonodromic tau function is shown to be equal to twist operator correlator. Our result is specialized to matrix monodromy data with quasi-permutation monodromy matrices and generalize in this case the relation between twist operator correlator and isomonodromic tau function to generic 2d CFTs using universal arguments.
\end{itemize}

\subsection{Remaining questions and future directions}
\begin{itemize}
    \item \textit{Explicit evaluation of twist operator correlator/tau function.} While our expression for stress-tensor one-point function holds for generic branched covers with non-abelian monodromy, it is not explicit enough for direct evaluation of twist operator correlator using stress-tensor method as the uniformizing map in general is not known explicitly beyond the cyclic cases; this is also the case in the path integral method where the coefficients of covering map are in general not known explicitly as functions of branch locus. However, the relation with tau function might provide a promising reformulation in light of the recent development in the CFT/isomonodromy correspondence~\cite{Gamayun:2012ma,Iorgov:2014vla,Gavrylenko:2018ckn}, which relates isomonodromic tau function associated with rank $N$ matrix Fuchsian equation to Fourier-transformed $W_N$ conformal blocks. The relation is made precise for generic $N=2$ (Virasoro) cases but only certain semi-degenerate cases for arbitrary $N$ due to technicalities in $W_N$ conformal blocks and therefore the relation doesn't immediately apply for the isomonodromic tau function related to twist operator correlator.\footnote{The semi-degeneracy essentially means that all but two of the spectra of monodromy matrices have $N-1$ degeneracies. While at first sight it seems that quasi-permutation matrices corresponding to transpositions satisfy this condition, the extra minus sign (needed for a cycle of even length~\cite{korotkin2003solution}) makes it actually have $N-2$ degeneracy.} A precise realization of the CFT/isomonodromy correspondence at arbitrary $N$ for generic matrix monodromy data would give explicit evaluation of (the universal Weyl anomaly part of) twist operator correlator via its relation with isomonodromic tau function.

    \item \textit{Generalization to higher genus $g \geq 2$.} A similar understanding, as in the genus zero and one case, of the general structure of twist operator correlator associated with generic branched covers with genus $g \geq 2$ for generic 2d CFTs has remained lacking in literature: in symmetric product orbifold context, most of the discussions have been focused on the lower genus cases motivated by their relevance for large $\Norb$ limit; in the context of replica trick calculation of quantum-information quantities, usually a single block domination prescription, valid for large $c$ 2d CFT, is used to calculate twist operator correlator at higher genus~\cite{Hartman:2013mia,Dutta:2019gen}; there are also results on free theories~\cite{Calabrese:2009ez,Dixon:1986qv,Zamolodchikov:1987ae}. While an explicit evaluation of twist operator correlator for generic 2d CFTs associated with generic branched covers is, as already in genus zero and one case, likely out of reach, one might hope to understand better its general structure such as i) the suitable formulation of path integral and stress-tensor method for generic 2d CFTs and their consistency in $g \geq 2$ cases and ii) the relation with tau function on Hurwitz space and isomonodromic tau function, which are indeed still well-defined at $g \geq 2$~\cite{kokotov2003taufunction,korotkin2003solution}.
\end{itemize}

\acknowledgments
We would like to thank Veronika Hubeny and Mukund Rangamani for related discussions, and Mukund Rangamani for comments on a draft of the paper. We would also like to thank Dmitry Korotkin for related correspondence. This work is partially supported by funds from the University of California.

\appendix
\section{Theta function conventions}
\label{sec-theta-function}
The genus $g$ Riemann theta function is defined as
\begin{equation}
    \Theta(\vb{u}|\vb*{\tau}) = \sum_{\vb{n} \in \mathbb{Z}^g} e^{2 \pi i \vb{n} \vdot \vb{u}} e^{\pi i \vb{n} \vdot \vb*{\tau} \vdot \vb{n}},
\end{equation}
and the Riemann theta function with characteristics $\vb{c} = \frac{\vb{a}}{2} + \frac{\vb*{\tau} \vdot \vb{b}}{2}$ can be defined as
\begin{equation}
    \Theta_{\vb{c}}\prn{\vb{u}|\vb*{\tau}} = \Theta \mqty[\vb{a} \\ \vb{b}]\prn{\vb{u}|\vb*{\tau}} = \exp \pi i \prn{ \frac{\vb{a} \vdot \vb*{\tau} \vdot \vb{a}}{4}  + \vb{a} \vdot \vb{u} + \frac{\vb{a} \vdot \vb{b}}{2}} \Theta\prn{\vb{u} + \vb{c} | \vb*{\tau}}.
\end{equation}
A characteristics $\vb{c} = \frac{\vb{a}}{2} + \frac{\vb*{\tau} \vdot \vb{b}}{2}$ is called half-integer if $\vb{a}, \vb{b} \in \mathbb{Z}^g$ and even/odd if $\vb{a} \vdot \vb{b}$ is even/odd.

At $g=1$, the Jacobi theta functions are related to Riemann theta function by
\begin{align}
    \theta_1(u|\tau) &= -\theta \mqty[1\\1]\prn{u|\tau}, \quad 
    \theta_2(u|\tau) = \theta \mqty[1\\0]\prn{u|\tau}, \nonumber\\
    \theta_3(u|\tau) &= \theta \mqty[0\\0]\prn{u|\tau}, \quad
    \theta_4(u|\tau) = \theta \mqty[0\\1]\prn{u|\tau} 
\end{align}
with $\theta_1(u|\tau)$ being the one with odd characteristics and others with even characteristics. Theta functions with argument $u=0$ are abbreviated as 
\begin{equation}
    \theta_\nu(0|\tau) = \theta_\nu(\tau), \quad \nu=1, \dots, 4.
\end{equation}

\bibliographystyle{JHEP}
\bibliography{refs}

\end{document}